\documentclass[conference,a4paper]{IEEEtran}
\usepackage{caption}
\usepackage{cite}
\usepackage{amsmath,amssymb,amsfonts}
\usepackage{algorithm}
\usepackage{algorithmic}
\usepackage{graphicx}
\usepackage{textcomp}
\usepackage{threeparttable}
\usepackage{xcolor}
\usepackage{xspace}
\usepackage{hyperref}
\usepackage{tikz}

\def\BibTeX{{\rm B\kern-.05em{\sc i\kern-.025em b}\kern-.08em
		T\kern-.1667em\lower.7ex\hbox{E}\kern-.125emX}}
\begin{document}
	
	\title{Increasing Resilience of Smart Home Agents
		\\
		
		\thanks{Identify applicable funding agency here. If none, delete this.}
	}
	
	\author{
		
		\IEEEauthorblockN{Christopher Terrazas, Eduardo Cotilla-Sanchez}
		\IEEEauthorblockA{\textit{School of Electrical Engineering and Computer Science} \\
			\textit{Oregon State University}\\
			Corvallis, Oregon, USA \\
			\{terrazac, ecs\}@oregonstate.edu}
		
	}
	
	\newcommand{\agentname}{Reflexion\xspace}
	\newcommand{\react}{ReAct\xspace}
	\newcommand{\reflexion}{Reflexion\xspace}
	\newcommand{\homeassistant}{HomeAssistant\xspace}
	\newcommand{\sage}{Sage\xspace}
	\newcommand{\sasha}{Sasha\xspace}
	\newcommand{\dspy}{DsPy\xspace}
	\newcommand{\vllm}{vLLM\xspace}
	\newcommand{\homebench}{HomeBench\xspace}
	\newcommand{\todocite}{ \textcolor{red}{[CITE]} }
	
\usetikzlibrary{positioning,arrows.meta,shapes.geometric,fit,backgrounds,calc}
	
	\maketitle

    \begin{tikzpicture}[remember picture,overlay]
  \node[anchor=south, yshift=15mm, text width=\textwidth,
        align=justify, font=\footnotesize]
    at (current page.south)
    {Accepted for publication at the 2026 International Conference on
     Smart Energy Systems and Technologies (SEST).\\
     \copyright~2026 IEEE. Personal use of this material is permitted.
     Permission from IEEE must be obtained for all other uses, in any
     current or future media, including reprinting/republishing this
     material for advertising or promotional purposes, creating new
     collective works, for resale or redistribution to servers or lists,
     or reuse of any copyrighted component of this work in other works.};
\end{tikzpicture}
	
	\begin{abstract}		
	Smart homes and smart devices are becoming more prevalent across millions of homes around the world. With the rise
	of AI, the smart home industry is quickly increasing its integration to
	manage common smart home tasks. However, existing work
	in large language models (LLMs) as agents within smart
	homes have shown minimal resilience due to limited environment
	scenarios or poor performance in complex tasks. We explore
	several strategies for LLMs as agents within the popular open-source software (OSS) smart home automation framework
	\homeassistant to increase overall smart home resilience. Our
	approach combines traditional supervised learning techniques
	and optimized prompting as the core learning process. We use
	a diverse set of LLMs covering different levels of reasoning
	and costs in our optimization pipeline and evaluate their
	performance on a subset of the hardest tasks in a smart
	home benchmark. We include ReAct and Reflexion agentic
	paradigms and reveal how both provide marginal return on
	investment compared to fine-tuned LLMs for multi-device
	control within \homeassistant but show promise in resilience tasks such as failure response.
	\end{abstract}
	
	\begin{IEEEkeywords}
		Agents, Edge Resilience, IoT, LLMs, Smart Home
	\end{IEEEkeywords}
	
	\section{Introduction}
		
	The smart home automation industry spans both commercial
	and OSS solutions. One of the issues with the latter is the user
	adoption of the software has a high learning curve and requires
	an in-depth level of technical expertise for setup and ongoing
	maintenance. This results in unsustainable use or total user
	abandonment for non-technical users. The burden extends to the developers of smart home frameworks due to its support of many diverse smart home devices \cite{b1}. LLMs as agents can help
	minimize the burden of smart home operational tasks, such
	as orchestrating smart device communication for a particular
	outcome. These types of tasks are synonymous with decision-
	making in the literature, where an agent decides an action
	within a particular environment given some current state.
	
	There are a vast number of day-to-day tasks that an OSS
	smart home user can perform within their smart home. These
	include actions such as device control and interaction, powering on and off smart lights, analyzing grid power input and output usage, or adjusting comfort devices such as heat pumps and HVACs. In smart home environments, it is critical to not only accomplish a task correctly, but to also avoid creating mishaps, especially in high-critical spaces like smart buildings within the medical space, defense complexes, or computational research facilities.
	
	Another common issue OSS smart home users face is troubleshooting. This process is similar to finding and triaging bugs within the field of software engineering. Troubleshooting for OSS smart home users consists of identifying problematic integrations, automations, or device conditions. For a new user, this can become overwhelming and difficult to sustain. For the OSS smart home user, the troubleshooting steps involve clicking their user interface to determine device status changes and deciphering error logs to pinpoint failure points. Internally, this is equivalent to the function of internal API endpoints within an OSS smart home framework. With the rapid improvements in popular LLMs, LLM-powered agents are more capable and can help a smart home user with troubleshooting tasks such as log parsing or reducing the speed of software development for engineers \cite{b2}.
	
	LLMs as agents have the ability to reason about their actions in their environment which make them suitable for exploration in the smart home space and as a tool for smart home users to increase their overall reliability and operational efficiency of their smart home.
	
	\section{Environments}
	
	\subsection{Home Assistant}\label{HomeAssistant}

	\homeassistant \cite{b3} has become the defacto OSS framework for smart home automation. It is written and maintained in Python, a popular programming language, that allows many hobbyists and programmers in the community to contribute to its development in a distributed fashion. However, HomeAssistant's modular architecture and technical documentation can become overwhelming for both new and experienced users. There are many ways to install the software to support: mini computers, docker containers, and many different operating systems and hardware architecture combinations. After installing the software, users can begin to add devices and write automations.
	
	We use \homeassistant as a basis for testing and deploying our agents due to its stable nature in development and its design to separate varying levels of control. For example, \homeassistant has authentication out of the box for users. This is a standard requirement for most smart home users and critical to restrict unauthorized access for smart homes integrated with LLMs and autonomous agents. The overall design of \homeassistant consists of a hierarchical format starting with their Home Assistant Operating System, then a Supervisor for managing the operating system, and finally the core software. For our resilience testing, we focus on the core software and its API endpoints to perform common smart home task failure responses.
	
	\section{Related Work}

	\subsection{Smart Home Agents}\label{Portability}
	
	A common strategy for interacting with LLMs is the chain of thought \cite{b4} process where a user and LLM make progress through guided prompts by the user. In both \sage \cite{b5} and \sasha \cite{b6}, prompts are  used to train the agents. In \sasha, their approach focuses on action plans that rely on semantic context to make progress towards ambiguous goals from the user. In \sage, prompts are dynamically generated to guide the LLM throughout the action space. 
	
	Instead of creating methods to influence LLM behavior by constructing prompts as an emphasis, we adopt the Demonstrate-Search-Predict framework's (\dspy) focus on programming through structured signatures and its built-in optimizers to preserve the agent's training process and let \dspy handle the prompt optimization. We find this approach less sensitive to changes on the application layer and supports multiple functionalities similar to LangChain \cite{b7} within \sage.  LangChain is an LLM framework that focuses on manual prompting and targets out of the box LLM development, but we chose DsPy because of its focus to support lower-level customization of LLM behaviors through modules. Another reason we chose to incorporate \dspy into our pipeline is because of the potential to optimize with existing machine learning techniques. For example, we configured our Reflexion agent with a custom heuristic for smart home tasks that we found to be pivotal in its performance.
	
	\sasha and \sage both demonstrate portability in their approach by testing, designing, and evaluating different LLMs within their software to demonstrate LLM's use of their particular technique. This is a crucial aspect to the scientific process, but it is also important for usability. For example, a user might switch LLMs within their smart homes or perhaps switch to a new smart home environment. We aim to take this step further and extend this flexibility to both the environment the smart home agent will operate in and the different scenarios within that smart home environment. In the latter, we focus on "deploying" these smart home agents within actual smart home environments in a way that is controlled and safe. For example, we discuss the guard rails we explored in \autoref{deployment}. We aim to slowly progress towards a production state where a smart home agent is actively managing a smart home environment within \homeassistant. This is important as there can be high-stakes devices and automations within these smart home environments. In our experiment, we accomplish a hybrid version within a controlled environment with a combination of real devices and fake devices that have been configured for use for our agents.

	\section{Agent LLMs and Testing}
	
	The setup for our experiments consists of prompt optimizing, transfer learning, evaluation, and deployment phases. The training and testing pipeline produces the agents we evaluate, and we further deploy these agents in a live \homeassistant environment. Our focus is to demonstrate the performance of the LLMs while minimizing manual tasks and costs per token for each agent.

	\subsection{Large Language Models}\label{sec:llms}
	
	We combine a mix of state of the art LLMs from both private industry and OSS. The cost for using private LLMs, especially agents, is significant \cite{b8}. To reduce the cost to train and test agents, we selected models that were cost efficient relative to industry pricing, but also performed well in reasoning tasks. To broaden the impact of our experiments, we chose to incorporate popular OSS LLMs as a core LLM engine in our experiments. For OSS models, we targeted medium to large LLMs because smaller models are broadly understood to struggle with agentic tasks \cite{b9}.
	
	From private industry, we used GPT-4 \cite{b10} as our teacher for bootstrapping. We then used GPT-5-mini  \cite{b11}, DeepSeek-R1 \cite{b12}, and Grok-4-fast-reasoning  \cite{b13}. From OSS models, we selected Qwen2.5 72B  \cite{b14}, Gemma-2-27B  \cite{b15}, and GPT-OSS-120b  \cite{b16}.
	
	\subsection{Benchmarks}
	
	\subsubsection{HomeBench} We implement the \homebench \cite{b17} benchmark which consists of  175{,}000 smart home test cases across five different categories: Valid single-device instruction, Invalid single-device instruction, Valid multi-device instruction, Invalid multi-device instruction, and Mix multi-device instruction (MM). In an effort to focus on complex tasks which agents are well-suited for, we selected the last category of multi-device instruction as the foundation for our analysis. We chose this category as it was the  most difficult in \homebench for all LLMs tested with a $38\%$ success score for the best performing LLM. These types of tasks involve the agent interacting with multiple devices and tools to accomplish a specific expected output.

	Once we identified the worst-performing category to test our agents, we extracted all of the 4{,}414 MM test cases. In each test case, there is an id, input, and expected output field. From the 4{,}414 MM test cases, we extracted $441$ test cases at random as our initial training and validation datasets. We extended the original input field and added context for each training data entry, similar to  what an agent would see during the agent decision-making phase in \homebench. For example, the agent sees the agent system instructions, the specific task device lists and tools available along with their argument signatures, and finally the entire user request. We construct all of this as the training data for the agent to reduce the planning iteration space for an agent by learning from an example of what it would see in this specific smart home environment. We then use this data to begin our supervised learning strategy to increase overall performance of all agents we test. The subset of data we created captures a diverse subset of the hardest category in the \homebench dataset. The size of the dataset also reduces the cost of running experiments significantly. 
	
	\subsubsection{Resilience Dataset}
	
	In an effort to increase resilience of agents in real smart home environments, we created a small resilience benchmark focused on failure recovery. Prior work has focused on common smart home tasks, but it does not prioritize more complex tasks. For example, there are smart home tasks that are more difficult to complete such as recovering failed automations, offline integrations, or monitoring and responding to abnormal behavior such as power spikes caused by the devices themselves or external factors such as power outages. The size of our resilience dataset is 50 test cases aiming to test our LLMs as agents' response to more difficult failure scenarios, which we hold 25 for training and testing. We used Claude Opus 4.7 \cite{b18} to generate our suite of failure response tests and manually verified each for correctness. The dataset is constructed among 10 distinct categories covering various scenarios a smart home user may encounter:  power failure, integration outage, automation partial execution, cascading dependency, pre-event hardening (such as natural disasters), life safety, cloud outage, sensor disagreement, network partition, and update rollbacks.

	\section{LLM Strategies}

	\subsection{Base}
	
	For the base LLM strategy, we use each of the LLMs in \ref{sec:llms} and then load that specific model and run our evaluation with no modifications to the LLM. For example, this is equivalent to integrating a third-party LLM provider via user credentials and feeding queries to the LLM through the HTTP protocol via APIs. We chose to incorporate this as an initial baseline strategy among our various reasoning levels of LLMs within the smart home domain because of the common use among the smart home community. 
	
	For example, there are various third-party cloud LLM providers that HomeAssistant users have installed as an integration to use wtihin their smart home. The base LLM strategy requires only creating and connecting the necessary credentials for each model provider. For a smart home user, they can install a particular cloud LLM integration and then provide these credentials to begin using the LLM within their HomeAssistant environment.
	
	\subsection{Optimized}
	
	The optimized strategy focuses on using DsPy's internal optimizers for a supervised learning approach to help all of our LLMs learn more about the smart home domain through training data. We explain and mention how we split this data within HomeBench and out resilience dataset in \ref{sec:train}. The optimized strategy is used to explore if using training data and additional context for an LLM can improve the smart home agent in various tasks. We chose to incorporate this LLM strategy due to the significant amount of smart home data provided within HomeBench.
	
	\subsection{ReAct}
	
	The ReAct paradigm interleaves an LLM within several distinct steps as its agent loop: reasoning, acting, and observing. It does this by using past traces of its actions and reasoning to help steer itself to a desirable outcome on its current task. The overall process is decomposed in an iterative manner, per decision iteration, so that the LLM can determine a grounded action based on its current trajectory and context.
	
	In smart homes, this might be a common task such as finding a specific smart vacuum within a smart home and changing its status. This sounds trivial, but there may be several smart vacuums docked across different floors, and finding the correct one is the challenging part and not necessarily changing its status. In other more complex smart home tasks, such as troubleshooting, the LLM's observations drive its reasoning and actions to increase progress on a particular task. In this context, the action might be to change a smart device status, stop a problematic smart home automation that is currently running, or create a complex automation request from a user orchestrating several smart devices within the smart home.
	
	We incorporate this popular agentic paradigm and focus its closed-loop reasoning system on the hardest smart home tasks within HomeBench and against our resilience dataset. When the ReAct agent receives a command from a user, it commences its feedback loop to accomplish these smart home tasks. We also chose to include ReAct to extend from the base and optimized LLM strategies, which focus accomplishing each smart home task in one attempt with no continuous reasoning or in a \textit{zero-shot} manner.

	\subsection{Reflexion}
	
	To explore diverse types of agents, we implemented the  \reflexion \cite{b19} agent as shown in Algorithm~\ref{alg:reflexionalgo}. This technique incorporates verbal reinforcement learning combined with self-reflections and long-term memory to assist the agent in complex tasks. The advantage to this approach is agent creators are able to utilize reinforcement learning strategies without the high compute costs of developing and training reinforcement learning models. The verbal portion acts as linguistic feedback to the agent on how it is doing in its current task episode based on the length of experiences stored in long-term memory. This is similar to how humans improve in difficult tasks by noting past failed experiences and adjusting to accomplish a particular goal. The authors showed that the \reflexion learning framework performed better than \react in complex tasks in the AlfWorld dataset \cite{b20}. 
	
	The internal architecture for \reflexion is agnostic to the agent it uses. For example it implements the \react algorithm \cite{b21} for the self-improvement LLM that generates actions called the Actor as shown in Figure 1. We implemented this algorithm in a custom \dspy module. Within this module, we implemented the evaluation and self-reflection portions of \reflexion combined with the long-term memory for the Actor (the LLM). For our Actor, we kept the \react portion and combined it with \reflexion's self-reflecting process. For both the self-reflection and evaluator, we used \dspy's Predict module to dynamically classify our desired output at each phase in the \reflexion algorithm. 
	
	In the self-reflection, we adopted the \reflexion authors' fixed parameter size of three for the total experience length per episode to minimize the token context burden commonly faced in LLMs within agent tasks. In the evaluation portion, we implemented a metric inspired from the \reflexion Alfworld test case that returns a scalar reward signal $r$ to the agent.  During the evaluation portion, the Evaluator, another LLM created with \dspy, looks at the agent's current trajectory thus far and checks for areas such as repeated actions contributing to the same errors. This portion is domain-specific for the agent's environment.
	
	\begin{figure}[t]
		\centering
		\begin{tikzpicture}[
			node distance=0.5cm and 0.6cm,
			box/.style={rectangle, draw, rounded corners=2pt, minimum height=0.7cm, minimum width=1.6cm, font=\small, align=center},
			actor/.style={box, fill=blue!10},
			eval/.style={box, fill=orange!15},
			refl/.style={box, fill=green!15},
			mem/.style={box, fill=gray!15, minimum width=1.4cm},
			traj/.style={box, fill=yellow!15, minimum width=1.8cm},
			env/.style={box, fill=red!10, minimum width=1.8cm},
			arr/.style={-{Stealth[length=2mm]}, semithick},
			dspy/.style={dashed, draw=purple!70, rounded corners=4pt, inner sep=5pt}
			]
			
			\node[env] (env) {HomeAssistant};
			
			\node[actor, right=1.2cm of env] (actor) {Actor\\(\react)};
			
			\node[traj, below=of actor] (traj) {Trajectory $\mathcal{H}$};
			
			\node[eval, right=of actor] (eval) {Evaluator};
			
			\node[refl, below=of eval] (refl) {Self-\\Reflection};
			
			\node[mem, below=of refl] (mem) {LTM\\($|M|{\leq}3$)};
			
			\draw[arr] (actor.west) -- node[above, font=\scriptsize] {$a_i$} (env.east);
			\draw[arr] (env.south) |- node[pos=0.75, above, font=\scriptsize] {$o_i$} (traj.west);
			
			\draw[arr] (actor.south) -- node[right, font=\scriptsize] {$\text{thought}_i$} (traj.north);
			
			\draw[arr] (traj.east) -- ++(0.2,0) |- (eval.west);
			
			\draw[arr] (eval) -- node[right, font=\scriptsize] {$r{=}0$} (refl);
			
			\draw[arr] (refl) -- node[right, font=\scriptsize] {$\sigma_i$} (mem);
			
			\draw[arr] (mem.south) -- ++(0,-0.35) -| (traj.south);
			
			\begin{scope}[on background layer]
				\node[dspy, fit=(actor)(eval)(refl)(mem)(traj),
				label={[purple!70, font=\scriptsize\itshape]above:Custom \dspy Reflexion Module}] {};
			\end{scope}
			
		\end{tikzpicture}
		\caption{\agentname architecture. The Actor (\react) interacts with HomeAssistant and writes to a shared trajectory $\mathcal{H}$. After each action, the Evaluator scores $\mathcal{H}$; on a zero reward, the Self-Reflection module produces a verbal plan $\sigma_i$ that is appended to a bounded FIFO long-term memory ($|M|\leq 3$). The contents of $M$ are appended back into the trajectory at every step, making prior reflections visible to the Actor on subsequent decisions within the same episode. All four components are implemented as custom \dspy modules.}
		\label{fig:architecture}
	\end{figure}
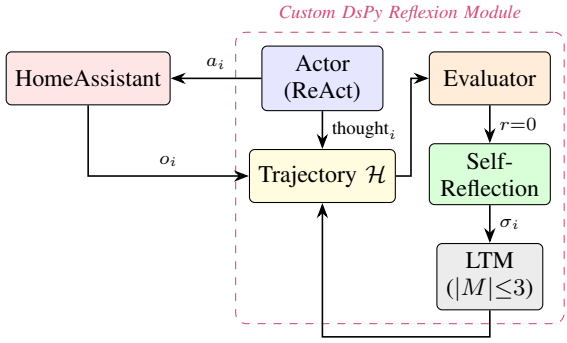

	\begin{algorithm}[t]
		\caption{\agentname: Inline Verbal-RL Loop for Smart Home Agents}
		\label{alg:reflexionalgo}
		\begin{algorithmic}[1]
			\REQUIRE Task inputs $x$, tools $\mathcal{T} \cup \{\textsc{finish}\}$, max iterations $N$
			\STATE Initialize trajectory $\mathcal{H} \gets \emptyset$
			\STATE Initialize long-term memory $M \gets \emptyset$ \COMMENT{bounded FIFO, $|M| \leq 3$}
			\STATE Initialize cumulative reward $r \gets 1$
			\FOR{$i = 1$ to $N$}
			\STATE $(\text{thought}_i, a_i, \text{args}_i) \gets \text{Actor}_{\react}(x, \mathcal{H})$
			\STATE $o_i \gets \text{Execute}(a_i, \text{args}_i)$ \COMMENT{Home Assistant API call}
			\STATE Append $(\text{thought}_i, a_i, \text{args}_i, o_i)$ to $\mathcal{H}$
			\STATE $r \gets r \cdot \text{Evaluator}(\mathcal{H})$ \COMMENT{$\in \{0, 1\}$; checks confused thoughts, repeated errors}
			\IF{$r = 0$}
			\STATE $\sigma_i \gets \text{SelfReflect}(r, \mathcal{H}, M)$ \COMMENT{verbal plan revising the strategy}
			\STATE $M \gets \text{Append}(M, \sigma_i)$ \COMMENT{evicts oldest if $|M| = 3$}
			\ENDIF
			\STATE Append $M$ to $\mathcal{H}$ as $\text{experience}_i$ \COMMENT{visible to Actor on next step}
			\IF{$a_i = \textsc{finish}$}
			\STATE \textbf{break}
			\ENDIF
			\ENDFOR
			\STATE $y \gets \text{Extract}(\mathcal{H}, x)$ \COMMENT{ChainOfThought fallback to final outputs}
			\RETURN $(y, \mathcal{H})$
		\end{algorithmic}
	\end{algorithm}

	\subsection{Training}\label{sec:train}
	
	We benchmark several machine learning techniques to train and test the smart home agents powered by multiple large language models. One of the issues with implementing agents is the tokens generated are greater due to their agent loops. To minimize this burden, we conducted a focused experiment with optimization, supervised learning, and transfer learning techniques to boost agent performance. Note, we were able to fine-tune the OSS model weights with MLX \cite{b22} and showed signals of improvement, but we were unable to serve locally due to the size of the LLMs as explained in \ref{sec:testing}.

	\subsubsection{Optimization}
	
	After extraction of the training data, we use the Demonstrate-Search-Predict framework \cite{b23} using \dspy  to optimize our agents with this training data. These agents learn about the \homebench environment, which was motivated from \homeassistant and prior work. In this approach, we aim to increase the abilities of the agent prior to its interaction with unseen data as seen in traditional supervised learning strategies. This has the benefit of reducing the planning space or the time taken for the agent to learn the basics of its environment. The key portion here is we do not have to perform manual prompting modifications which are brittle to future agent modifications.
	
	We use \dspy's Multiprompt Instruction PRoposal Optimizer Version 2  (MIPROV2) \cite{b24} to automatically optimize the prompts and instructions for smart home tasks. We used DsPy’s built-in ReAct module and wrote a custom DsPy
	Reflexion module for our experiments.  We annotated our smart home agent modules with the appropriate docstrings to explain the domain of smart homes and the agent's responsibilities within the smart home environment.
	
	\subsubsection{Supervised Learning}
	
	For all of our experiments, we used \homebench as a foundation for our smart home agents. To reduce costs and focus our strategy on agent evaluation, we used statistical power analysis and tested on a subset of the 175{,}000 \homebench dataset. Our subset consisted of 4{,}414 smart home automation tasks from the worst performing category across the state of the art LLMs. 
	
	\subsubsection{Transfer Learning}
	
	From this subset, we took 441 random examples from the original dataset to use as training data to optimize our best model on this task. After prompt optimization with MIPRO, we use transfer learning by having our best model act as a teacher for our less-capable LLMs. We use this strategy to reduce the overall costs of our experiments and to show techniques to squeeze agentic performance from the best performing LLMs. 
	
	\subsection{Testing}\label{sec:testing}
	
	For the testing portion of our experiment, we sampled 250 random examples for a 95\% confidence interval  with a 6.2\% margin of error. We evaluated each LLM under all four strategies (Base, Opt, ReAct, Reflexion) using this sample, then wrote a Python script to invoke the \homebench test suite. For the private industry LLMs, we updated the configurations accordingly to include the relevant API keys and API base urls per the OpenAI standard \cite{b25}.
	
	For the OSS LLMs, we used OpenRouter \cite{b26}, an OSS LLM hosting provider for inference. We attempted to self-host with Ollama \cite{b27}, but ran into memory issues during inference. We explored other viable solutions such as \vllm \cite{b28}, a high-throughput and memory-efficient inference and serving engine for LLMs which might be more suitable for running larger models locally. However, our machine was not compatible with the \vllm software.

	\section{Deployment}\label{deployment}

	\subsection{Guardrails}
	
	To ensure the agent had limited access in its scope, we restricted its available tools to the HomeAssistant Rest API. This API consists of a set of endpoints designed to manage connected smart devices, integrations, and automations. HomeAssistant supports LLM interactions with the Home Assist API, but we chose not to use this because the scope was very limited and excluded all administrative tasks for the LLM. For example, an LLM is required to interact with this API through intents over exposed entities. To achieve finer-grain control over the aspects of the smart home and to give our agents greater autonomy, we chose our custom approach of integrating with the HomeAssistant Rest API.  For each tool, we also included additional context for the smart home agent through annotated docstrings that detailed crucial information such as the tool name, expected number of parameters, and the expected output. 
	
	For security, we created a temporary token that the agent must use for authorization for each of its tool endpoints. The token can also be revoked manually by an authorized user within the HomeAssistant settings dashboard before its expiration. Our goal here was to simulate a privileged user's access within the smart home, and the ability to revoke and grant access instantly for the smart home user. We did not give the agent full access to all of the HomeAssistant APIs including the Hypervisor to remove any interaction with the operating system and network configurations. This approach reduces the ability of the smart home agent to execute undesirable actions that might arise from hallucinations or prompt injections by manually writing and configuring its tools.
	
	For the deployment environment, we configured the HomeAssistant dashboard with a hybrid approach of dummy devices and real smart devices such as a smart light. However, we did not include any critical devices in the smart home to minimize the overall risk level of the agent and the samrt home devices themselves.
	
	\subsection{Agent Interaction}
	
	Once deployed, we were able to interact with the agent via code and feed it queries to fetch information such as the current weather from the weather integration, statuses of various smart home devices, and it was able to generate working automations written in YAML. We were able to feed the agent natural language commands to successfully turn on and off smart lights. We also tried to access unauthorized endpoints within the Hypervisor and were unable to access any of the restricted areas via the agent commands. However, during this process, the agent would attempt to fetch the desired information from our prompts with tools it did have access to, but could not.
	
	\section{Evaluation}

\begin{table}[htbp]
	\caption{F1 Evaluation on HomeBench MultiDevice Tasks (\%)}
	\begin{center}
		\begin{threeparttable}
			\begin{tabular}{|c|c|c|c|c|}
				\hline
				\textbf{Model}&\multicolumn{4}{|c|}{\textbf{LLM Strategy}} \\
				\cline{2-5}
				\textbf{Name} & \textbf{\textit{Base}}& \textbf{\textit{Opt}}& \textbf{\textit{ReAct}}&\textbf{\textit{Reflexion}} \\
				\hline
				gpt-oss-120b & 86.7 & 88.6 & 85.4 & 85.0 \\
				gpt-5-mini   & 86.2 & 86.2 & 85.3 & 84.7 \\
				grok-4-1     & 86.3 & 89.9 & 88.0 & 87.8 \\
				deepseek-r1  & 79.0 & 85.4 & 77.7 & 76.6 \\
				qwen-2.5-72b & 46.4 & 52.4 & 45.4 & 44.6 \\
				gemma-2-27b  & 34.3 & 37.7 & 27.7 & 28.8 \\
				\hline
			\end{tabular}
			\begin{tablenotes}
				\footnotesize
				\item Sample-averaged F1 over $n{=}250$ scenarios sampled from 4{,}000
				(95\% confidence, $\pm6.2\%$ worst-case margin).
			\end{tablenotes}
		\end{threeparttable}
	\end{center}
	\label{tab1}
\end{table}

   We compute the F1 score similarly to HomeBench and compare predicted commands with a gold standard, then we calculate the unweighted mean across F1 scores for all of our examples. Our results show that optimized prompting and zero-shot LLMs are a good foundation for accomplishing complicated multi-device smart home tasks as seen in	Table 1. For most of the LLMs, the addition of agents degraded their performance compared to their base and opt level performance. The greatest disparity we observed was gemma-2-27b, with a drop in F1 score by 19.2\% in ReAct and a 16\% drop in Reflexion. This further supports that smaller models struggle in agentic reasoning loops. 
	
	However, it is worth noting that the other LLMs showed minimal improvements by adding ReAct and Reflexion agents. The only LLM to show an improvement from its base performance in both ReAct and Reflexion was grok-4-1 with a marginal 2\% increase.

\begin{table}[htbp]
	\caption{F1 Evaluation on Resilience Dataset (\%)}
	\begin{center}
		\setlength{\tabcolsep}{3pt}
		\begin{threeparttable}
			\begin{tabular}{|c|c|c|c|c|}
				\hline
				\textbf{Model}&\multicolumn{4}{|c|}{\textbf{LLM Strategy}} \\
				\cline{2-5}
				\textbf{Name} & \textbf{\textit{Base}}& \textbf{\textit{Opt}}& \textbf{\textit{ReAct}}&\textbf{\textit{Reflexion}} \\
				\hline
				gpt-oss-120b & 43.2 $\pm$ 8.2 & 45.2 $\pm$ 8.8 & 42.3 $\pm$ 8.7  & 47.1 $\pm$ 10.2 \\
				gpt-5-mini   & 43.8 $\pm$ 8.6 & 45.0 $\pm$ 8.6 & 44.7 $\pm$ 7.9  & 44.4 $\pm$ 8.8 \\
				grok-4-1     & 47.0 $\pm$ 9.6 & 47.7 $\pm$ 9.3 & 48.6 $\pm$ 10.0 & 46.1 $\pm$ 9.4 \\
				deepseek-r1     & 43.7 $\pm$ 8.4 & 45.0 $\pm$ 9.1 & 46.6 $\pm$ 9.1 & 41.1 $\pm$ 8.3 \\
				qwen-2.5-72b & 41.2 $\pm$ 8.0 & 41.2 $\pm$ 8.0 & 41.3 $\pm$ 7.9  & 41.6 $\pm$ 8.4 \\
				gemma-2-27b  & 41.2 $\pm$ 7.5 & 35.3 $\pm$ 5.3 & 37.6 $\pm$ 5.1  & 33.2 $\pm$ 7.0 \\
				\hline
			\end{tabular}
			\begin{tablenotes}
				\footnotesize
				\item Values are mean F1 $\pm$ 95\% confidence t-interval ($n{=}25$).
			\end{tablenotes}
		\end{threeparttable}
	\end{center}
	\label{tab2}
\end{table}

		For our resilience dataset, we see a different pattern emerge where agents are generally performing as good or better than zero-shot LLMs as seen in Table 2.

	\section{Discussion and Conclusion}\label{discussion}
	
	We showed \react and Reflexion's performance compared to several different optimized LLMs on a subset of the \homebench dataset. Our results show that optimized prompting and zero-shot LLMs are a good foundation for accomplishing complicated multi-device smart home tasks. In an effort to better leverage agents in smart homes, we created a small resilience benchmark where agentic strategies as a whole outperformed  baseline LLMs on resilience tasks. This was opposite of the trend we saw in multi-device smart home tasks, where agents under-performed for most models. In an actual smart home environment, we recommend a hybrid approach where zero-shot LLMs can handle multi-device smart home tasks and have agents for failure response tasks. 
	
	Future work can explore the creation of more complex smart home benchmarks where agentic paradigms may perform better  as in our failure response focus. Furthermore, exploration is needed in the reduction of overall training and computation costs of interacting with agents to minimize the token consumption of the planning process. For agents, this cost grows by a considerable factor compared
	to zero-shot LLMs. We recommend techniques similar to our strategy of using transfer learning and optimization techniques to boost overall agent performance. With this strategy, we were able to push OSS LLMs performance in the \homebench's
	most difficult domain by a significant margin. Our approach can be used to reduce inference costs further by self-hosting a fine-tuned OSS LLM, which many HomeAssistant users opt in for \cite{b29}. The self-hosting approach also increases smart home resilience by removing the dependence on the internet for full
	smart home agent operation. We acknowledge the difficulty and sustainability of this specific approach as it requires constructing a dataset for specific domain tasks. However, small datasets can be instrumental in improving performance
	using zero-shot techniques.

\end{document}